\documentclass[titlepage]{article}

\pdfoutput=1

\usepackage[left=1.2in,right=1in,top=1in,bottom=.8in]{geometry}
\usepackage{graphicx}
\usepackage{multicol}
\usepackage{caption}
\usepackage{subcaption}
\usepackage{float}
\usepackage{amsmath}
\usepackage{enumitem}
\usepackage{cite}
\DeclareMathOperator{\arccosh}{arccosh}

\newtheorem{theorem}{Theorem}
\newtheorem{corollary}{Corollary}

\begin{document}

\title{Rendering Non-Euclidean Geometry in Real-Time Using Spherical and Hyperbolic Trigonometry}
\author{Daniil Osudin \and Dr Chris Child \\ \\City, University of London \and Prof Yang-Hui He}
\date{08/06/2019}

\maketitle

\tableofcontents

\newpage

\begin{figure}[htp]
\centering
\begin{subfigure}{.33\textwidth}
  \centering
  \includegraphics[width=.95\linewidth]{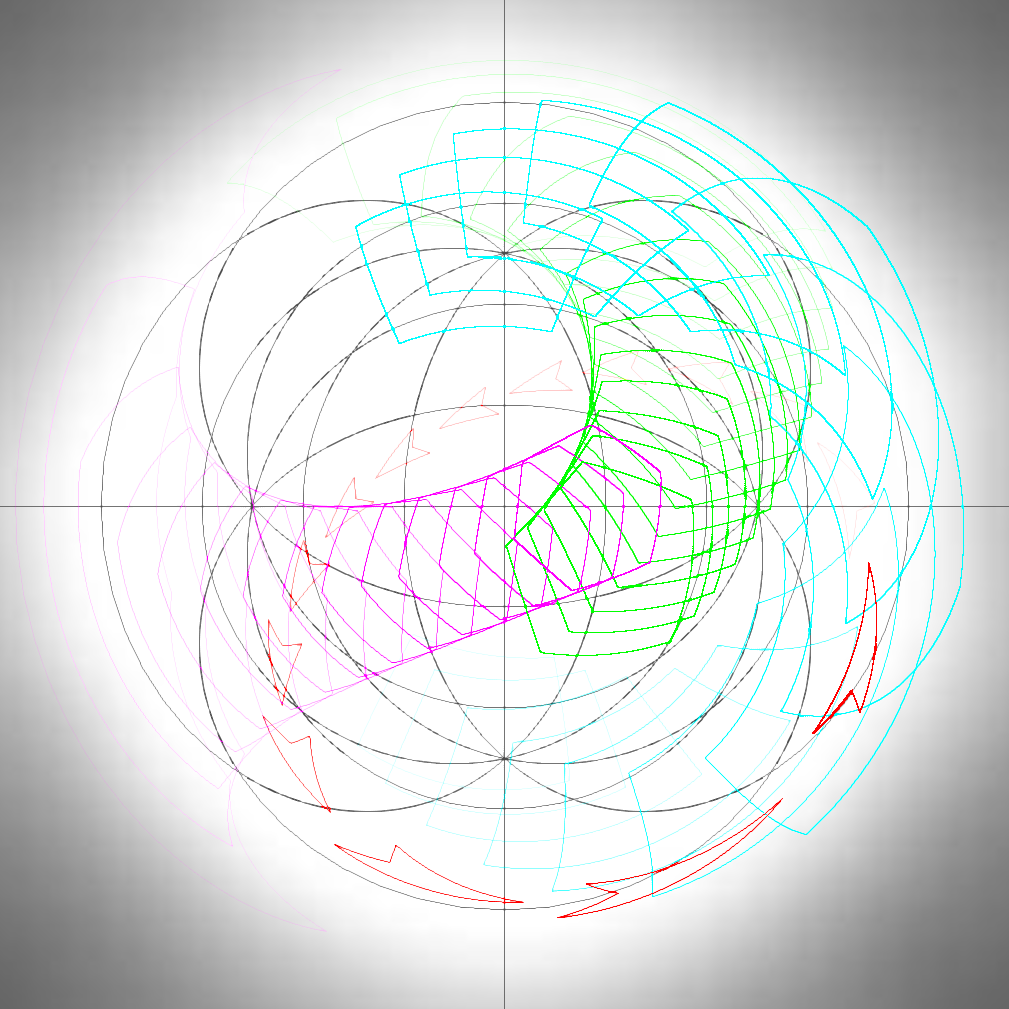}
  \caption{Spherical space timelapse}
  \label{fig:VertexCoordinates1}
\end{subfigure}%
\begin{subfigure}{.33\textwidth}
  \centering
  \includegraphics[width=.95\linewidth]{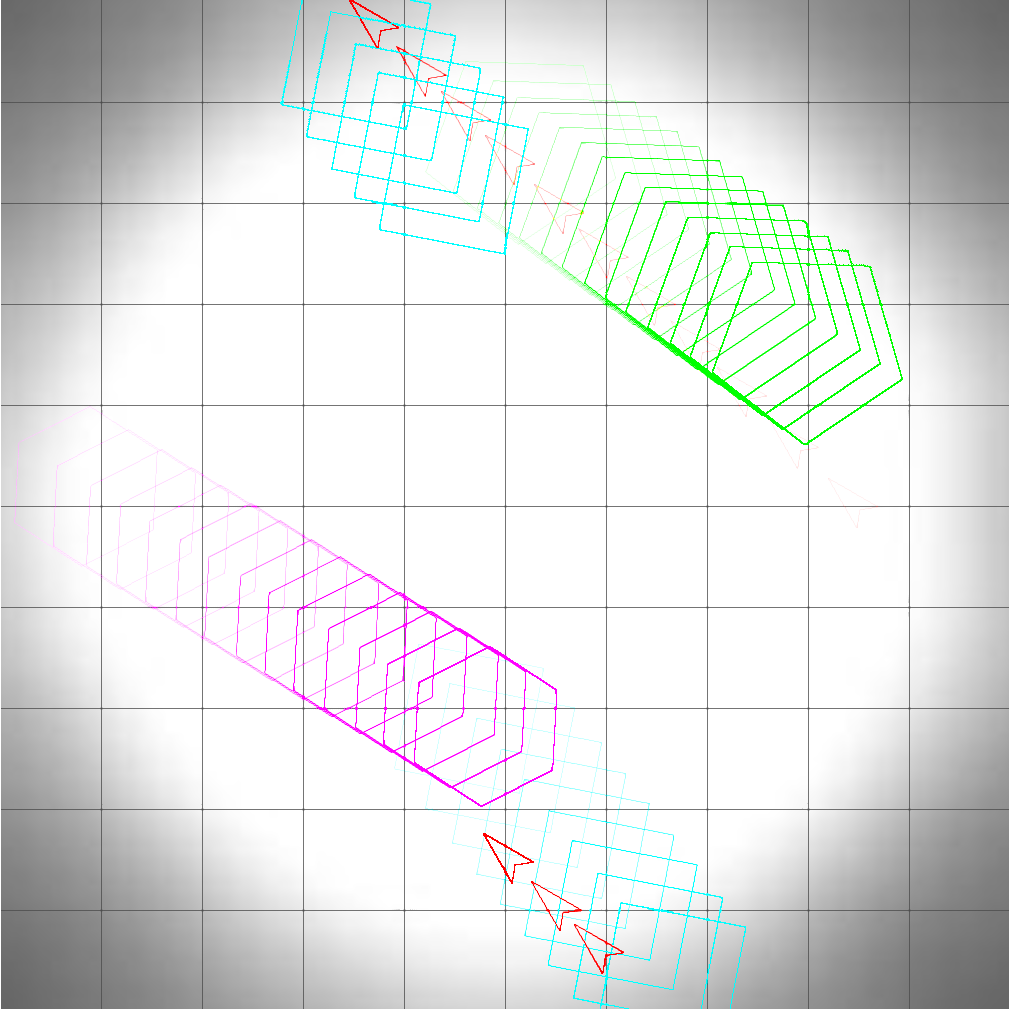}
  \caption{Spherical space timelapse}
  \label{fig:VertexCoordinates1}
\end{subfigure}%
\begin{subfigure}{.33\textwidth}
  \centering
  \includegraphics[width=.95\linewidth]{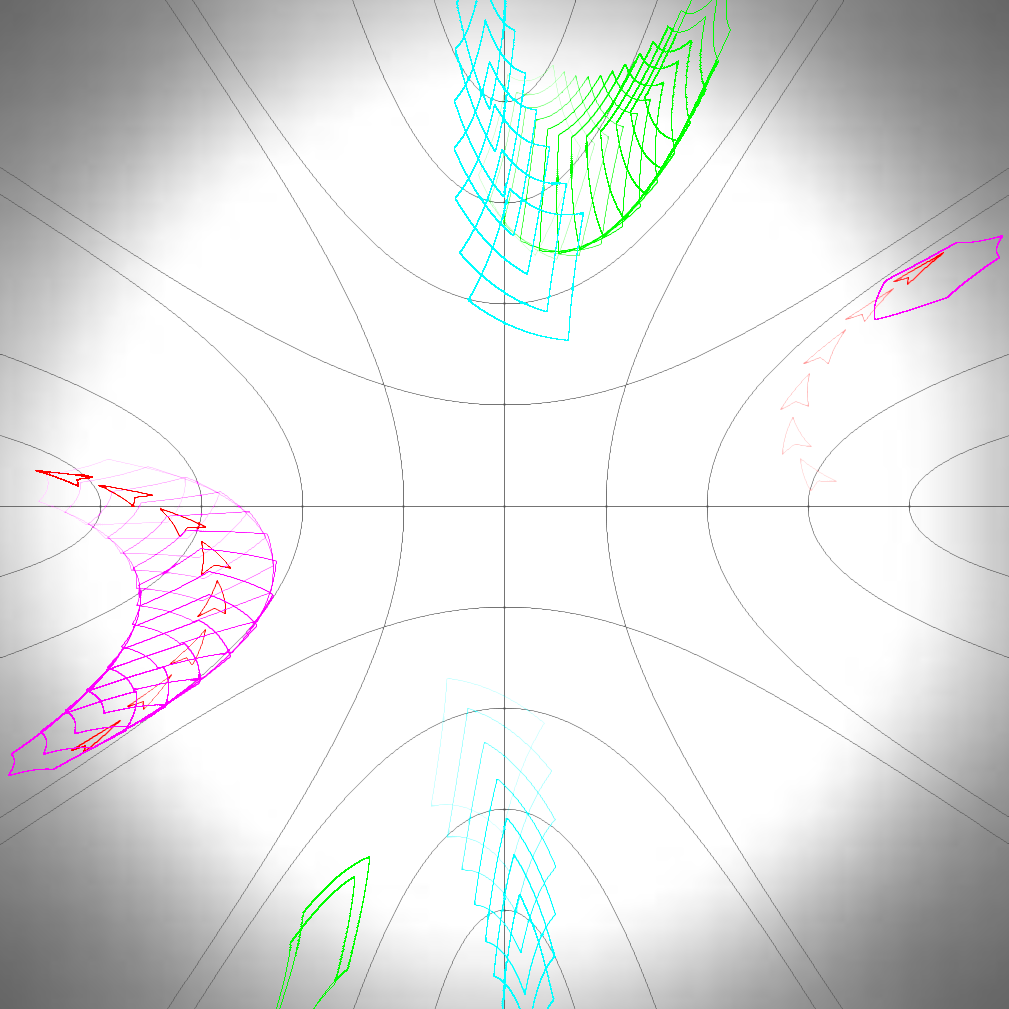}
  \caption{Hyperbolic space timelapse}
  \label{fig:VertexCoordinates2}
\end{subfigure}
\raggedright
\vspace{-5pt}
\caption{Time-lapse images of multiple objects moving through spherical (a), planar (b) and hyperbolic (c) 2D space calculated and rendered by the physics/graphics engine described below}
\vspace{-6pt}
\end{figure}

\begin{multicols}{2}

\section{Abstract}
This paper introduces a method of calculating and rendering shapes in a non-Euclidean 2D space. In order to achieve this, we developed a physics and graphics engine that uses hyperbolic trigonometry to calculate and subsequently render the shapes in a 2D space of constant negative or positive curvature in real-time. We have chosen to use polar coordinates to record the parameters of the objects as well as an azimuthal equidistant projection to render the space onto the screen because of the multiple useful properties they have. For example, polar coordinate system works well with trigonometric calculations, due to the distance from the reference point (analogous to origin in Cartesian coordinates) being one of the coordinates by definition. Azimuthal equidistant projection is not a typical projection, used for neither spherical nor hyperbolic space, however one of the main features of our engine relies on it: changing the curvature of the world in real-time without stopping the execution of the application in order to re-calculate the world. This is due to the projection properties that work identically for both spherical and hyperbolic space, as can be seen in the Figure 1 above. We will also be looking at the complexity analysis of this method as well as renderings that the engine produces. Finally we will be discussing the limitations and possible applications of the created engine as well as potential improvements of the described method.
\vspace{-10pt}
\section{Introduction}
Non-Euclidean geometry is a broad subject that takes its origin from Euclid's work Elements \cite{heath:1956}, where he defined his five postulates. This field encompasses any geometry that arises from either changing the parallel postulate (Euclid's fifth postulate) or the metric requirement. In this study we will be focussing solely on traditional non-Euclidean 2D geometries: Spherical geometry and Hyperbolic geometry, illustrated on Figure 2 (a) and (c) respectively.

\begin{figure}[H]
	\vspace{-10pt}
	\centering
	\includegraphics[trim = 3.25cm 22.25cm 3.25cm 3.25cm, clip, width=.495\textwidth]{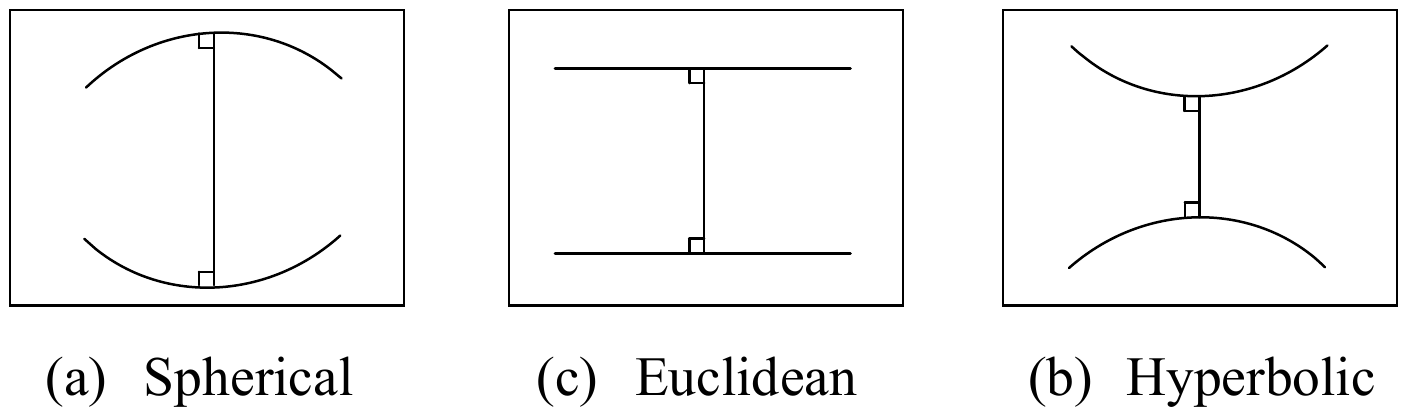}
	\vspace{-20pt}
	\caption{Comparison of parallel lines in the 2D spaces of different curvature}
	\label{fig:Noneuclid}
	\vspace{-10pt}
\end{figure}

In spherical geometry all geodesics (straight lines in a non-planar space) intersect, so there are no parallel lines. Even if the lines start parallel, they don't preserve the same distance along their length and instead appear to `bend' towards each other. In fact any two great circles will intersect twice (unless they are one and the same). Spherical geometry is used in multiple fields: navigation, GPS, architecture and aerospace engineering among others. However in most circumstances the calculations are done on a surface of a 3D sphere instead of the 2D spherical plane, and rendering if any uses the orthographic projection of the sphere. Figure 3 below, for example, illustrates Earth's eastern hemisphere:

\begin{figure}[H]
	\vspace{-10pt}
	\centering
	\includegraphics[width=.35\textwidth]{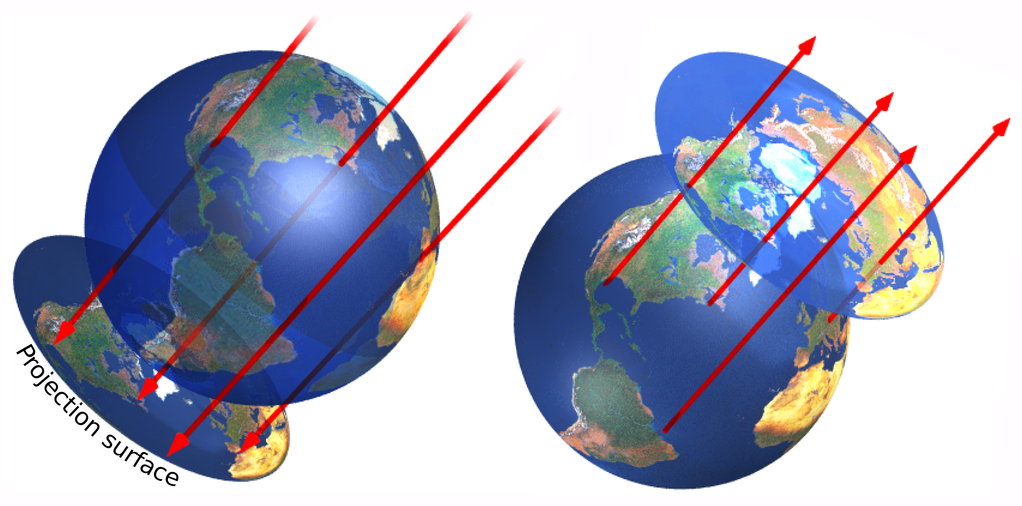}
	\caption{Orthographic projection of the sphere\cite{furuti:2012}}
	\label{fig:orthographicSphere}
	\vspace{-10pt}
\end{figure}

In cartography, multiple other projections are used and one of them has been chosen by us to be a focus of this study: Azimuthal equidistant projection. A detailed explanation of this projection and its advantages for this model is described in the Method section.

In Hyperbolic geometry, any line can have an infinite number of parallel lines, as the lines appear to `bend' away from each other. Elliptic geometry is more tangible and intuitive than hyperbolic geometry, due to people interacting with it more and the possibility of a 2D elliptic plane to be embedded into a 3D space. Hyperbolic geometry is more abstract; however it is used heavily in mathematics, astrophysics and theoretical physics, particularly for calculations involving general relativity. The usual projection used to represent hyperbolic geometry is the Poincare disc.

\begin{figure}[H]
	\vspace{-5pt}
	\centering
	\includegraphics[width=.2\textwidth]{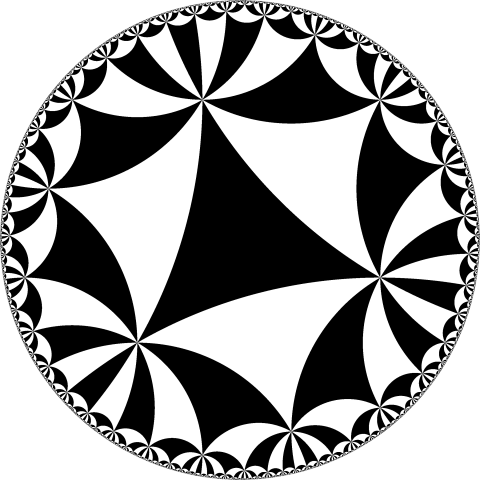}
	\caption{Poincare disc with hyperbolic tiling\cite{tamfang:2011}}
	\label{fig:poincarDisc}
	\vspace{-10pt}
\end{figure}

Our aim was to create a physics and graphics engine capable of calculating and rendering objects in a 2D space of arbitrary constant curvature. Our additional objective was to make it possible for the curvature to be modified in real-time during the execution of the engine. We approached this by developing a method of calculating the object's position as well as the vertices of its shape in polar coordinates using spherical \cite{todhunter:1886} or hyperbolic trigonometry \cite{carslaw:1916} \cite{traver:2014} and then rendering the objects onto the screen using an azimuthal equidistant projection. The real-time change of curvature was achieved by having the shapes of the object's dynamically recalculated in order to always use the current curvature variable.

Figure 1 above shows the time-lapse results of the engine rendering multiple objects in spherical (a), planar (b) and hyperbolic (c) geometries. The resulting engine can be used and built upon for multiple different purposes, such as creating video games or animations in non-Euclidean environment, help visualise the mathematics of non-Euclidean space and create tools for use in different areas of research.

\section{Method}

The method chosen is using hyperbolic and spherical trigonometry, as well as equidistant azimuthal projection and polar coordinates system in order to calculate object positions and render the objects in an arbitrary 2D space of constant positive or negative curvature.

\subsection{Polar coordinates}
\begin{figure}[H]
	\vspace{-5pt}
	\centering
	\includegraphics[trim = 2.5cm 23.3cm 12.2cm 1.5cm, clip, width=0.65\linewidth]{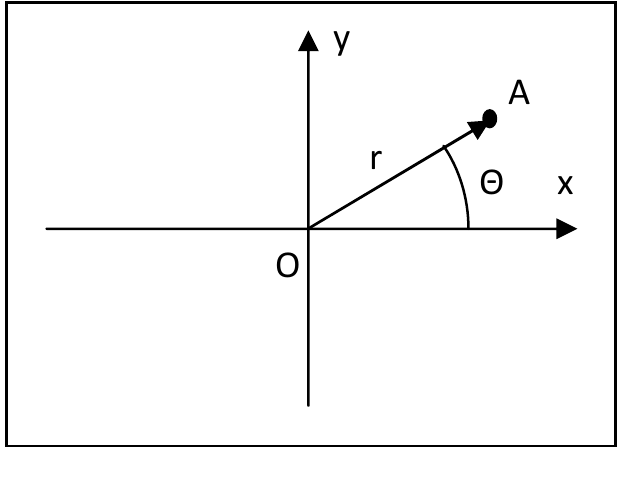}
	\vspace{-10pt}
	\caption{Position of a point A in polar coordinates. It has distance $r$ and bearing $\theta$ from the reference point $O$}
	\label{fig:PolarCoordinates}
	\vspace{-10pt}
\end{figure}

A polar coordinate system of the form $(r,\theta)$ is used in this model for all of the object positions and calculations instead of Cartesian coordinates.
The centre of the of the screen is taken as a reference point (analogous to the origin point in Cartesian coordinates) for the distance coordinate, $r$, while eastbound direction is set to be the reference direction for the bearing coordinate, $\theta$.

Cartesian coordinates need to be adapted in order to be used for constant negative and positive curvature spaces, as the parallel lines, which are essential to pinpoint the location in Cartesian space, are fundamentally different in a non-Euclidean space.

On the other hand, polar coordinates work just as well in any space of constant curvature, as the distance from a reference point and bearing from a reference direction still exist. This principle will help unify the coordinate system for any curvature of the space and ultimately allow the real-time change of curvature without any complications.

\subsection{Projection}

In order to render a non-Euclidean 2D space onto a flat screen, a projection has to be used which translate the points in a curved space onto a flat 2D space that is a screen.
\begin{figure}[H]
	\vspace{-25pt}
	\centering
	\includegraphics[trim = 2.2cm 19.2cm 11.3cm 4.8cm, clip, width=0.7\linewidth]{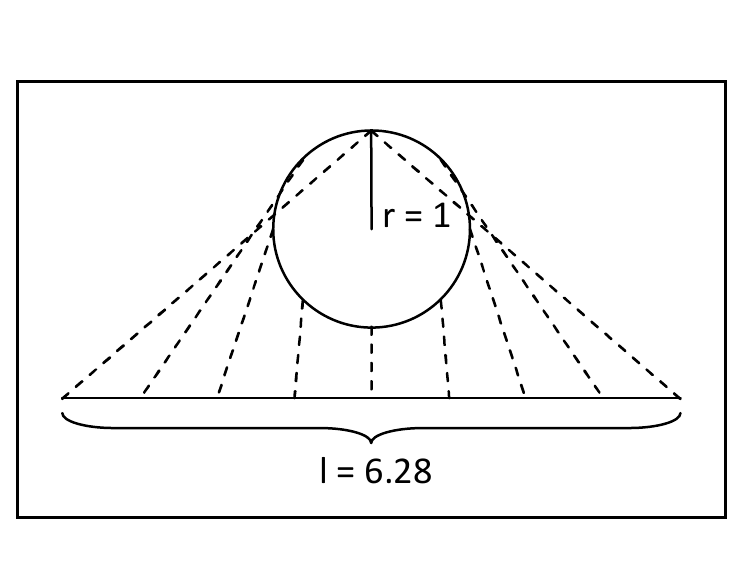}
	\vspace{-10pt}
	\caption{Side view of unfolding a sphere or radius $r$ onto a plane of size $l$ using Azimuthal Equidistant projection. The diagram is to scale.}
	\label{fig:SideProjection}
	\vspace{-10pt}
\end{figure}

We chose to use azimuthal equidistant projection, as this projection has useful properties that fit the purpose well. Firstly, by definition, distances and bearing from the centre of the projection are preserved. This fits perfectly with the Polar coordinates: for any point, its position vector on the projection is the position vector in the curved space. Secondly, this projection is intuitive, so makes the non-Euclidean geometry easier to understand. Finally, this projection can be used with no change to represent both spherical and hyperbolic 2D spaces.

\subsection{Mathematical analysis}
The calculations in this model are split into two parts: movement of the objects and rendering of the shapes, so the analysis will also be split into the corresponding subsections. Initially, it is important to overview how the model is structured and how the objects, shapes and the space are defined in the model. 

First, as previously mentioned, the model uses a polar coordinate system for all calculations and these are only converted into Cartesian coordinates for the final step of rendering, in order to utilise the rendering functions available through OpenGL. Hence, when a vector is mentioned in this section, we are referring to a Polar Vector in a form $(r,\theta)$. 

The screen (rendering space) is limited to a circle of an arbitrary size. When the object's centre moves past the circumference of the circle, it is repositioned to the antipodal point on the circle with the velocity preserved. This is implemented in order to keep the objects in the visible area on the screen. 

Each object has a set of parameters that contain its properties, essential for the calculations described in this section. 
\\
Object variables:
\begin{itemize}[topsep=-1pt]
  \setlength{\itemsep}{1pt}
  \setlength{\parskip}{0pt}
  \setlength{\parsep}{0pt}
\item Position vector
\item Velocity vector
\item Acceleration vector
\item Rotation angle
\item Rotation speed
\item Shape
\end{itemize}
Shape has a List of position vectors for each vertex. These vectors are defined in local coordinates with the reference point being the centre of the object (point defined by object's position vector in global coordinates) and reference direction is taken as the opposite of the object's position vector (illustrated on the Figure 8 below). This method simplifies the trigonometric calculations explained in the following subsections. All of the vertices are defined to be connected by the edges that lie on the geodesics which standardizes the definition of the object in a way that is independent of the world's curvature. As such, in order to have a shape with a curved edge (e.g. semicircle), multiple vertices will need to be set to represent the curved edge.

\vspace{-20pt}
\begin{figure}[H]
	\centering
	\includegraphics[trim = 2.4cm 21.6cm 10.8cm 1.95cm, clip, width=0.7\linewidth]{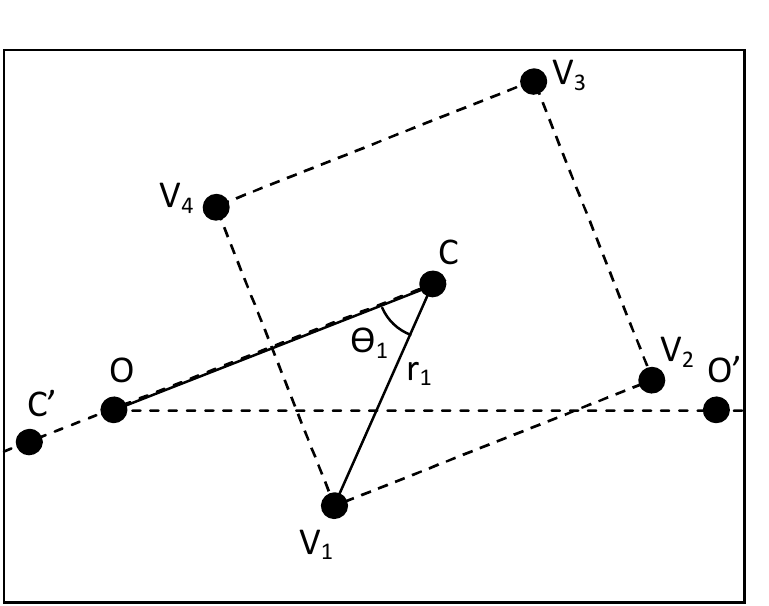}
	\vspace{-5pt}
	\caption{Local coordinate system of the object. Key: O, the reference point with coordinates $(0,0)$; C, position of the object and reference point of the local coordinate system with coordinates $(r_c, \theta_c)$; V$_1$, vertex of the illustrated object with local coordinates $(r_1,\theta_1)$; V$_1$, V$_2$, V$_3$, other vertices; OO', reference direction; CC', local reference direction}
	\label{fig:mixed}
\end{figure}
\vspace{-16pt}

\textbf{Note}: Diagrams below are shown in planar geometry in order to keep them clear, the triangle is still a good representation for the analogous spherical and hyperbolic triangles that arise depending on the world curvature setting. This is because all of the known values remain unchanged irrespective of curvature, as they are pre-set by definition of the shape and object in the engine.

\subsubsection{Rendering the shape}

In order to render the shape, global coordinates have to be calculated for all of the vertices of the shape. This model is using spherical and hyperbolic trigonometry in order to calculate these values.

Let $K \in [-1,1] \subset \Re$ s.t.

$K = 0 \Rightarrow$ Euclidean Geometry

$K > 0 \Rightarrow$ Spherical Geometry, $r = \frac{1}{\sqrt{K}}$ 

$K < 0 \Rightarrow$ Hyperbolic Geometry, $k = \frac{1}{\sqrt{K}}$

\begin{figure}[H]
	\vspace{-15pt}
	\centering
	\includegraphics[trim = 2.5cm 21.6cm 11.1cm 2.0cm, clip, width=0.9\linewidth]{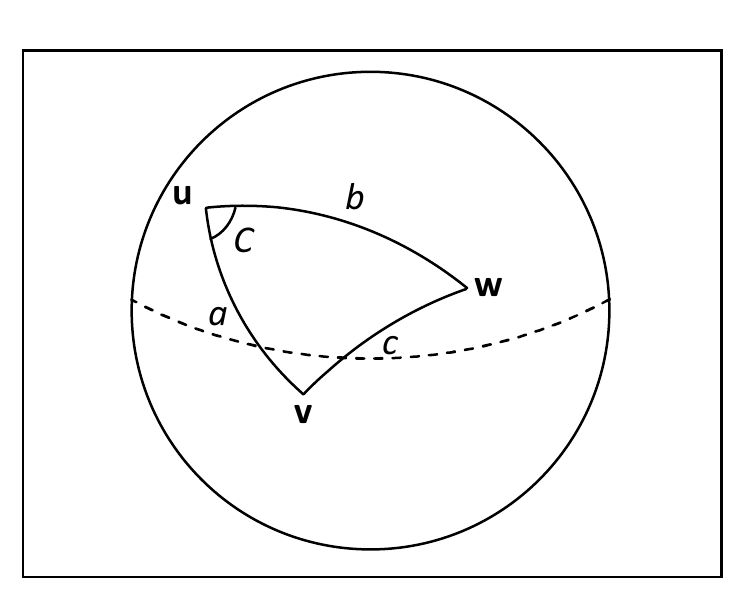}
	\vspace{-10pt}
	\caption{Spherical triangle on the surface of a sphere}
	\label{fig:SphericalTriangle}
	\vspace{-10pt}
\end{figure}

\begin{theorem}
For a sphere of radius \textbf{r} and hence Gaussian curvature $K = \frac{1}{r^2}$, as well as a spherical triangle on its surface described by points \textbf{u}, \textbf{v} and \textbf{w}, connected by great circles that form the edges \textit{a}, \textit{b}, \textit{c} (interpreted as subtended angles) and an angle \textit{C} (See figure~\ref{fig:SphericalTriangle}), the spherical law of cosines states: \cite{gellert:1989}
\vspace{-5pt}
\begin{equation} 
\cos{\frac{c}{r}} = \cos{\frac{a}{r}} \cos{\frac{b}{r}} + \sin{\frac{a}{r}} \sin{\frac{b}{r}} \cos{C}
\end{equation}
\vspace{-5pt}
\end{theorem}

\begin{figure}[H]
	\vspace{-15pt}
	\centering
	\includegraphics[trim = 2.5cm 21.6cm 11.1cm 2.0cm, clip, width=0.95\linewidth]{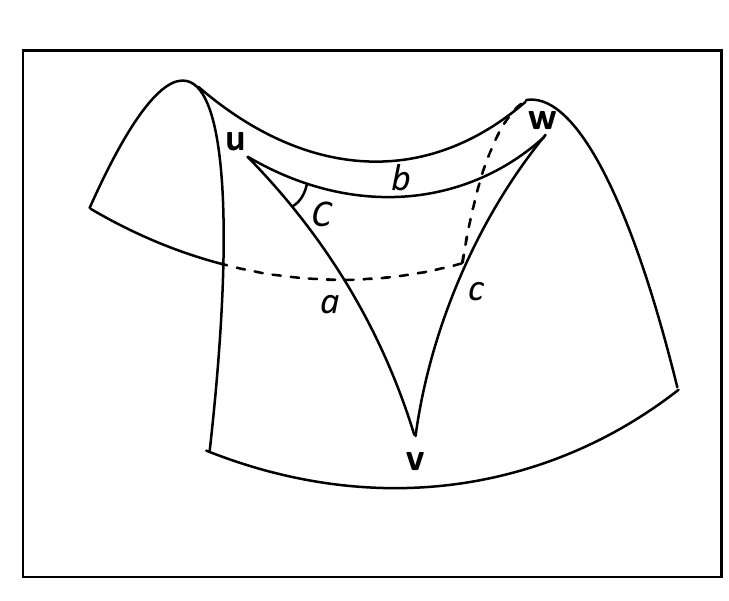}
	\vspace{-10pt}
	\caption{Hyperbolic triangle on the surface of a hyperbolic plane}
	\label{fig:HyperbolicTriangle}
	\vspace{-10pt}
\end{figure}

\begin{theorem}
For a hyperbolic plane with Gaussian Curvature $K = -\frac{1}{k^2}$ and a hyperbolic triangle on its surface described by points \textbf{u}, \textbf{v} and \textbf{w}, connected by geodesics that form the edges \textit{a}, \textit{b} and \textit{c}, as well as an angle \textit{C} (See figure~\ref{fig:HyperbolicTriangle}), the hyperbolic law of cosines states: \cite{gray:1979}
\vspace{-5pt}
\begin{equation} 
\cosh{\frac{c}{k}} = \cosh{\frac{a}{k}} \cosh{\frac{b}{k}} - \sinh{\frac{a}{k}} \sinh{\frac{b}{k}} \cos{C}
\end{equation}
\end{theorem}
\vspace{-5pt}

\end{multicols}
\vspace{-10pt}
\begin{figure}[H]
\centering
  \includegraphics[trim = 0.3cm 21.1cm 0.525cm 1.5cm, clip, width=.95\linewidth]{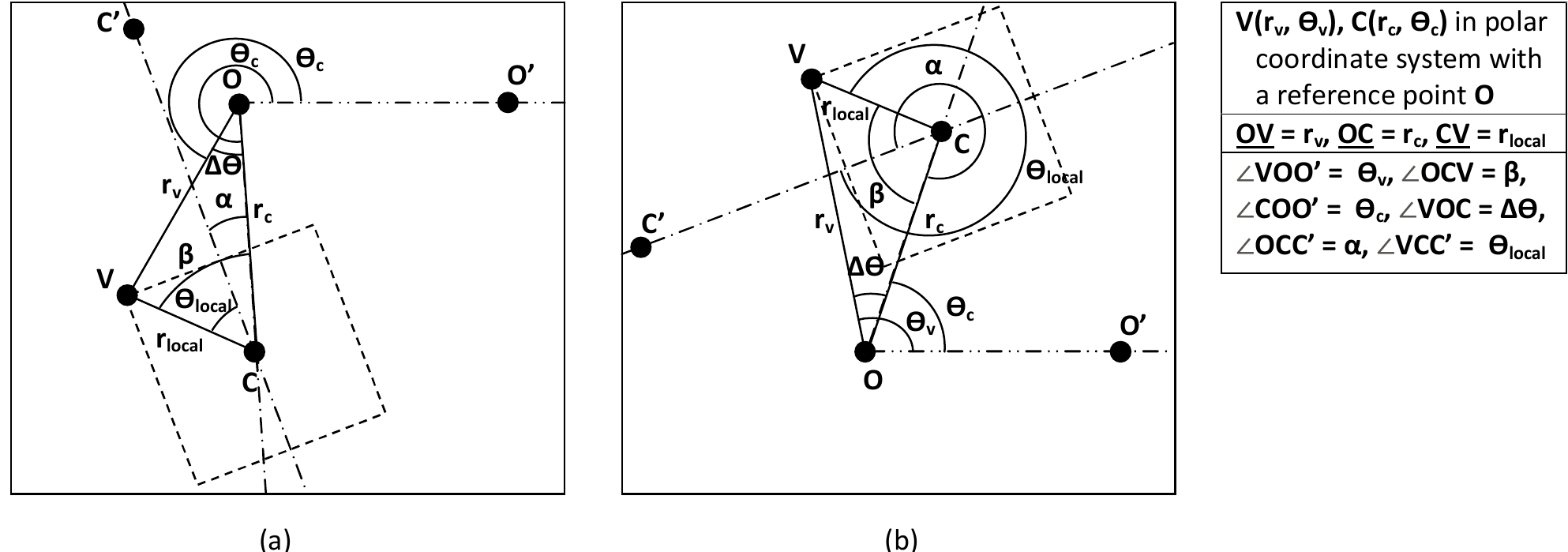}
  \label{fig:VertexCoordinates}
\raggedright
\vspace{-5pt}
\caption{Finding the $\theta$ and $r$ coordinates of an object's vertices through a hyperbolic/spherical triangle $OCV$. Key: O, reference point of the global polar coordinate system with coordinates $(0, 0)$; C, position of the object with coordinates $(r_c, \theta_c)$ and origin of the local polar coordinate system of the object; V, vertex of the object with global coordinates $(r, \theta)$ and local coordinates $(r_{local}, \theta_{local})$; $\alpha$, object's angle of rotation in local coordinates; $\beta$, angle between object position vector (line OC) and line from centre of the object to the vertex position (line CV). Case (a): Sum of $\theta_{local}$ and $\alpha$ is less than $\pi$; case (b): sum of $\theta_{local}$ and $\alpha$ is more than $\pi$}
\vspace{-10pt}
\end{figure}
\begin{multicols}{2}

\textbf{Note}: in order to simplify the equations below, all of the lengths will be divided by $r$ or $k$ depending on the value of K. Then the lengths that we are searching for will be multiplied by $r$ or $k$ to get the final answer.

\begin{corollary}
Given: \textbf{O$(0, 0)$, C$(r_c, \theta_c)$, V$(r_v, \theta_v)$, \underline{OC} = $r_c$, \underline{CV} = $r_{local}$, \angle COO' = $\theta_c$, \angle OCC' = $\alpha$, \angle VCC' = $\theta_{local}$}

Find: $r_v$, $\theta_v$ = ?

If $K > 0$, then:
\vspace{-5pt}
\begin{equation}
\begin{split}
r_v = \arccos{(\cos{ r_c} \cos{r_{local}}} + \sin{r_c}\\
\sin{r_{local}} \cos{\beta})
\end{split}
\end{equation}
\vspace{-5pt}
\begin{equation} 
\Delta \theta_v = \arccos{(\frac {\cos{r_{local}}-\cos{r_c} \cos {r_v}}{\sin{r_c} sin{r_v}})}
\end{equation}
\vspace{-5pt}
\\

If $K < 0$, then:
\vspace{-5pt}
\begin{equation} 
\begin{split}
r_v = \arccosh{(\cosh{ r_c} \cosh{r_{local}} - \sinh{r_c}} \\
\sinh{r_{local}} \cos{\beta)}
\end{split}
\end{equation}
\vspace{-5pt}
\begin{equation}
\Delta \theta_v = \arccos{(\frac {\cosh{r_c} \cosh{r_v}- \cosh{r_{local}}}{\sinh{r_c} \sinh{r_v}})}
\end{equation}
\vspace{-5pt}

In order to find $r_v$, we first find \angle OCV = $\beta$. $\beta = \alpha + \theta_{local}$; however if $\Pi < \beta < 2\Pi$, use the explementary angle of $\beta$ instead. This is done to determine to which side of \underline{OC} the triangle lies, which will be used to find $\theta_v$. In order to find $\theta_v$, we calculate \angle VOC = $\Delta\theta$, which is then added to or subtracted from $\theta_c$, depending on whether the explementary angle of $\beta$ was taken or not.
\end{corollary}

\end{multicols}
\vspace{-5pt}
\begin{figure}[H]
\centering
  \includegraphics[trim = 0.3cm 19.3cm 0.425cm 2.1cm, clip, width=.95\linewidth]{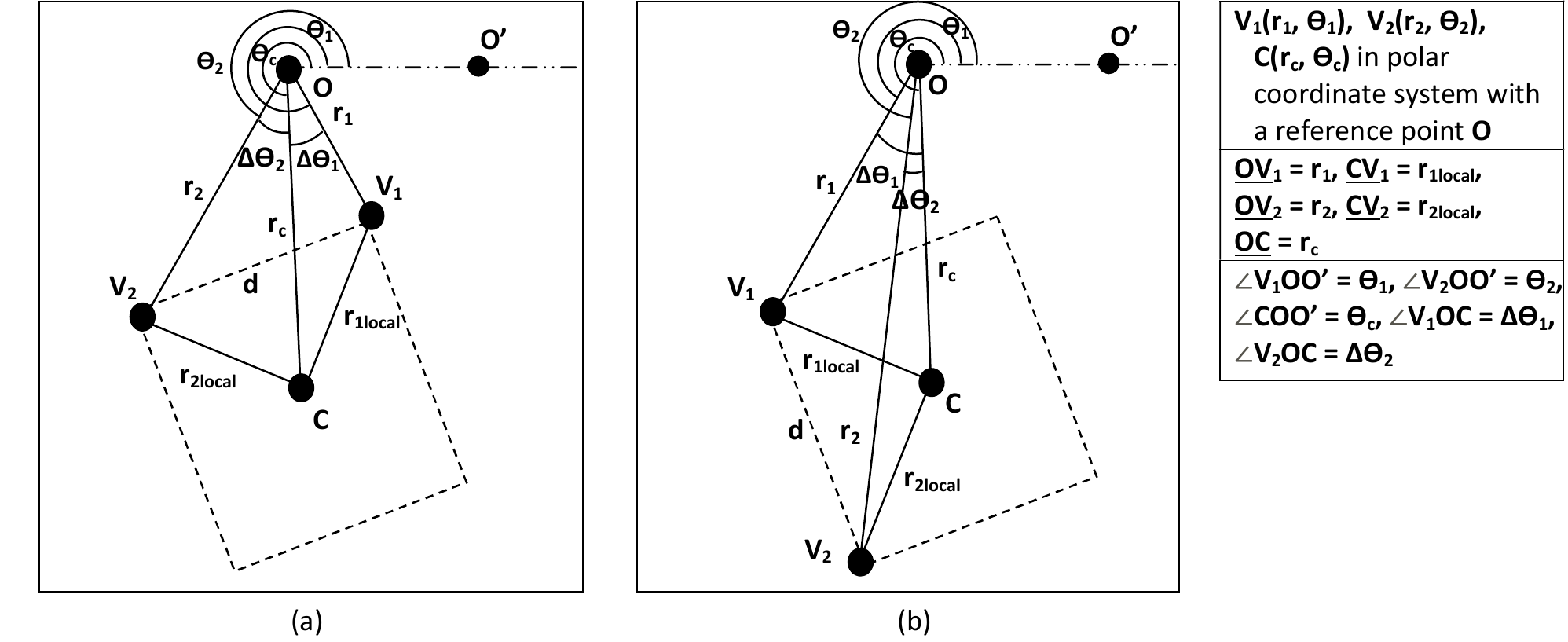}
  \label{fig:deltaTheta}
\raggedright
\vspace{-5pt}
\caption{Finding preliminaries to calculate intermediate points between two vertices: the length of the edge as well as the angle between the position vectors of the two vertices of this edge. Key: $O$, reference point of the global coordinate system; $C$, position of the object; $V_1, V_2$, vertices of the object; $d$, edge $V_1V_2$; $r_1, r_2$, r coordinates of the respective vertex; $\Delta\theta_1, \Delta\theta_2$, difference between $\theta$ coordinate of C and the respective vertex; $r_{1local}, r_{2local}$, r coordinates of the respective vertex in local coordinates. Overall $\Delta\theta$ is the angle between $r_1$ and $r_2$. And there are two cases to be considered. In the first case the angles $\Delta\theta_1 and \Delta\theta_2$ are diverging (a), so $\Delta\theta$ is the sum of these angles; the second case is when these angles converge, in which case $\Delta\theta$ is the absolute value of the difference between the two angles.}
\vspace{-10pt}
\end{figure}
\begin{multicols}{2}

\textbf{Note}: The above steps need to be repeated to find every vertex of an object. However that is not enough information to draw curved lines onto a screen. In order to proceed with rendering, a geodsic between the two vertices should be tessellated into smaller straight lines, which could be rendered on the screen. Hence intermediate points should be found along the lines connecting the vertices of the shape. To find them, we need to find the length of the edge between the two vetices as well as the total angle between the position vectors of the two vertices (illustrated on the figure 11 above).

\begin{corollary}
Given: \textbf{O$(0, 0)$, C$(r_c, \theta_c)$, V$_1(r_1, \theta_1)$, V$_2(r_2, \theta_2)$, \underline{OC} = $r_c$, \underline{OV$_1$} = $r_1$, \underline{OV$_2$} = $r_2$, \underline{CV$_1$} = $r_{1local}$, \underline{CV$_2$} = $r_{2local}$, \angle COO' = $\theta_c$, \angle V$_1$OO' = $\theta_1$,  \angle V$_2$OO' = $\theta_2$}

Find: $d$, $\Delta\theta$ = ?

In order to find $\Delta\theta$, we first find $\Delta\theta_1$ and  $\Delta\theta_2$:
\vspace{-5pt}
\begin{equation}
\Delta\theta_1 = \theta_c - \theta_1
\end{equation}
\vspace{-5pt}
\begin{equation}
\Delta\theta_2 = \theta_c - \theta_2
\end{equation}
\vspace{-5pt}

Here we could have 2 cases: diverging angles and converging angles (see diagram 10).

Angles diverge if $\Delta\theta_1 < 0$, $\Delta\theta_2 < 0$ or $\Delta\theta_1 > 0$, $\Delta\theta_2 > 0$. Then:
\vspace{-5pt}
\begin{equation}
\Delta\theta = \lVert \Delta\theta_1 \rVert + \lVert \Delta\theta_2 \rVert
\end{equation}
\vspace{-5pt}

Angles converge if $\Delta\theta_1 > 0$, $\Delta\theta_2 < 0$ or $\Delta\theta_1 < 0$, $\Delta\theta_2 > 0$. Then:
\vspace{-5pt}
\begin{equation}
\Delta\theta = \lVert \Delta\theta_1 - \Delta\theta_2 \rVert
\end{equation}
\vspace{-5pt}

In order to find $d$, consider $\triangle$\textbf{OV$_1$V$_2$}.

If $K > 0$, then:
\vspace{-5pt}
\begin{equation}
\begin{split}
d = \arccos{(\cos{ r_1} \cos{r_2} + \sin{r_1}} \\
\sin{r_2} \cos{\Delta \theta})
\end{split}
\end{equation}
\vspace{-5pt}

If $K < 0$, then:
\vspace{-5pt}
\begin{equation} 
\begin{split}
d = \arccosh{(\cosh {r_1} \cosh{r_2}– \sinh{r_1}} \\
\sinh{r_2} \cos{\Delta \theta})
\end{split}
\end{equation}
\vspace{-5pt}
\end{corollary}

\textbf{Note}: We also need to record which of $\Delta\theta_1$ and $\Delta\theta_2$ is the greater angle, as that determines the direction of the edge $d$ used in the next step.

\begin{figure}[H]
\vspace{-5pt}
\centering
\includegraphics[trim = 2.4cm 19.9cm 8.6cm 3.4cm, clip, width=1.0\linewidth]{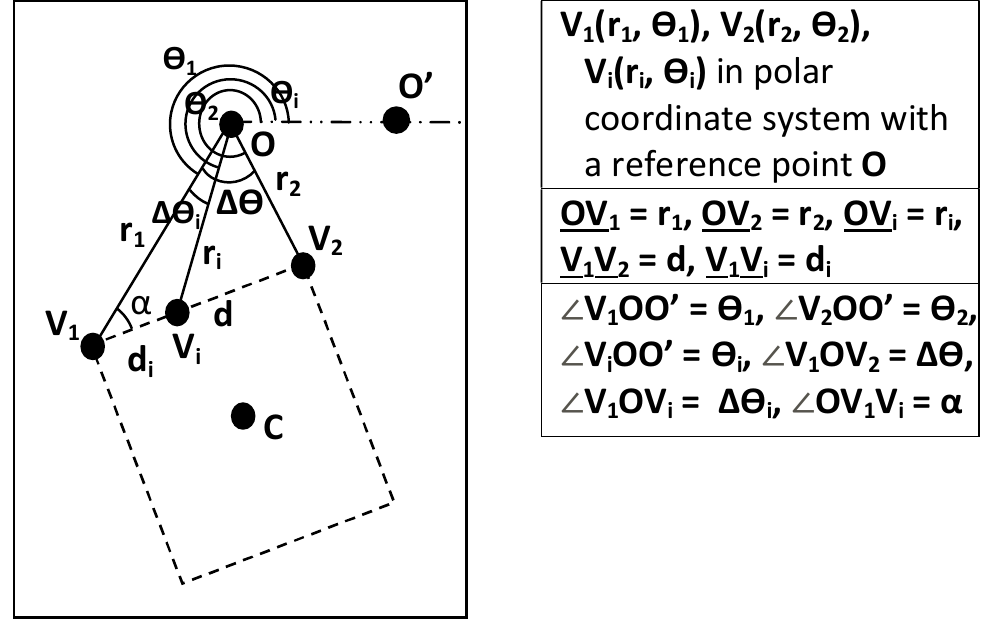}
\vspace{-20pt}
\caption{Finding Intermediate points between two vertices in order to render the edge. Key: $O$, reference point of the global coordinate system; $C$, position of the object; $V_1, V_2$, vertices of the object; $d$, edge $V_1V_2$; $V_i$, point on $d$; $r_1, r_2, r_i$, $r$ coordinates of the respective point; $\Delta\theta_i$, difference between $\theta$ coordinate of $V_1$ and $V_i$; $\Delta\theta$, angle between $OV_1$ and $OV_2$}
\label{fig:Spherical1}
\vspace{-10pt}
\end{figure}

\textbf{Note}: distance d is divided into a number of equal parts in order to find the distance $d_i$ for each of the points on the edge V$_1$V$_2$. The number of segments depends on the object tesselation vaiable.

\begin{corollary}
Given: \textbf{O$(0, 0)$, V$_1(r_1, \theta_1)$, V$_2(r_2, \theta_2)$, V$_i(r_i, \theta_i)$, \underline{OV$_1$} = $r_1$, \underline{OV$_2$} = $r_2$, \underline{V$_1$V$_2$} = d, \underline{V$_1$V$_i$} = $d_i$, \angle V$_1$OO' = $\theta_1$,  \angle V$_2$OO' = $\theta_2$, \angle V$_1$OV$_2$ = $\Delta\theta$}

Find: $r_i$, $\theta_i$ = ?

If $K > 0$, then:
\vspace{-5pt}
\begin{equation}
\alpha = \arccos{ (\frac{\cos{r_2}- \cos{r_1} \cos{d}}{\sin{r_1} \sin{d}})}
\end{equation}
\vspace{-5pt}
\begin{equation}
\begin{split}
r_i = \arccos{(\cos{r_1} \cos{d_i} + \sin{r_1}}\\
\sin{d_i} \cos{\alpha})
\end{split}
\end{equation}
\vspace{-5pt}
\begin{equation}
\Delta \theta_i = \arccos{(\frac{\cos{d_i}-\cos{r_1} \cos{r_i}}{\sin{r_1} \sin{r_i}})}
\end{equation}
\vspace{-5pt}

If $K < 0$, then:
\vspace{-5pt}
\begin{equation}
\alpha = \arccos{ (\frac{\cosh{r_1} \cosh{d}-\cosh{r_2}}{\sinh{r_1} \sinh{d}})}
\end{equation}
\vspace{-5pt}
\begin{equation}
\begin{split}
r_i = \arccosh{(\cosh{r_1} \cosh{d_i} – \sinh{r_1}}\\
\sinh{d_i} \cos{\alpha})
\end{split}
\end{equation}
\vspace{-5pt}
\begin{equation}
\Delta \theta_i = \arccos{(\frac{\cosh{r_1} \cosh{r_i} - \cosh{d_i}}{\sinh{r_1} \sinh{r_i}})}
\end{equation}
\vspace{-5pt}

Angle $\alpha$ is calculated as an inbetween step to find the angle opposite r$_i$ to apply the rule of cosines. Then $r_i$ and subsequently $\Delta \theta_i$ can be found using the cosine rule. To calculate the distance $r_i$, the triangle $V_1OV_i$ is considered. Using the previously found angle $\alpha$ as well as the known lengths $r_1$ and $V_1V_i$ ($d_i$) a hyperbolic/spherical cosine rule can be applied (illustrated on figure 12).

Then to find actual coordinates of the point $V_i$, $r_i$ should be multiplied by $r$ or $k$ depending on the value of K; $\Delta \theta_i$ should be added to or subtracted from angle $\theta$1, depending on the direction of the edge $d$, determined previously.
\end{corollary}

\subsubsection{Updating object position}
\end{multicols}
\begin{figure}[H]
\vspace{-15pt}
\centering
  \includegraphics[trim = 0.4cm 9.6cm 5.5cm 2.5cm, clip, width=.95\linewidth]{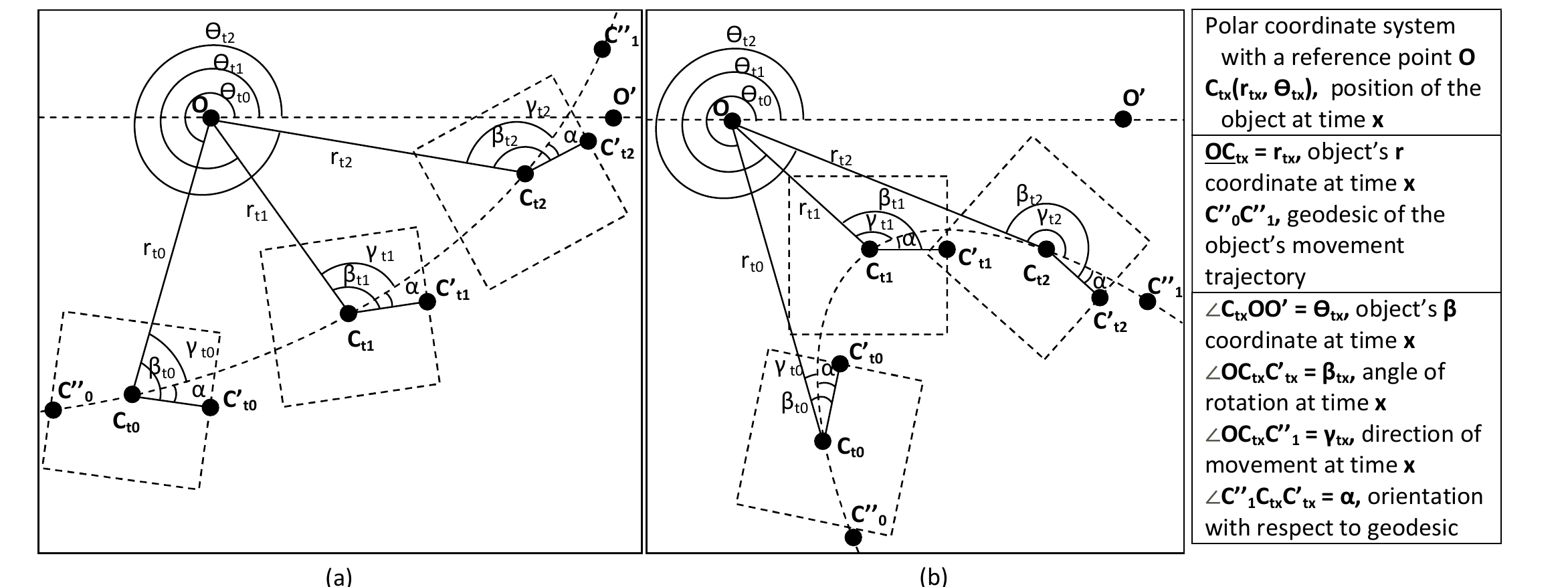}
  \label{fig:deltaTheta}
\raggedright
\vspace{-5pt}
\caption{Movement of the object along a hyperbolic in Spherical (a) and Hyperbolic (b) space. Orientation with respect to the geodesic is kept the same (angle $\alpha$ is constant) if the object is not rotating. Key: O, reference point of the global coordinate system with coordinates $(0, 0)$; C$_{tx}$, position of the object with coordinates $(r_tx, \theta_tx)$ and origin of the local coordinate system of the object at time $x$; $\beta_{tx}$, object's angle of rotation in local coordinates at time $x$; $\alpha$, angle between the geodesic and the object's local reference direction (should not change if the object doesn't rotate); $\gamma_{tx}$, direction of object's velocity vector}
\vspace{-10pt}
\end{figure}
\begin{multicols}{2}

\textbf{Note}: In addition to updating the global position vector of an object, the velocity vector and rotation angle have to be updated accordingly in order to keep their orientation towards the geodesic consistent (for a non rotating object; for a rotating object, extra rotation over time should be added after the position of the object was recalculated).

\begin{corollary}
Given: \textbf{O$(0, 0)$, C$_{t0}(r_{t0}, \theta_{t0})$, C$_{t1}(r_{t1}, \theta_{t1})$, \underline{OC$_{t0}$} = $r_{t0}$, \underline{C$_{t0}$C$_{t1}$} = $r_p$, \angle C$_{t0}$OO' = $\theta_{t0}$, \angle OC$_{t0}$C'' = $\gamma_{t0}$, \angle OC$_{t0}$C'$_{t0}$ = $\beta_{t0}$}

Find: $r_{t1}$, $\theta_{t1}$, $\gamma_{t1}$, $\beta_{t1}$ = ?

$\gamma_{t0}$ should be in the range $0$ to $\pi$, take explementary angle if $\alpha_0 > \pi$. This will determine the direction of the movement with respect to the reference point (needed to calculate $\theta_{t1}$).

Let \angle OC$_{t1}$C$_{t0}$ = $\gamma'_{t1}$

If $K > 0$, then:
\vspace{-5pt}
\begin{equation}
\begin{split}
r_{t1} = \arccos{\cos{r_{t0}} \cos{r_p}+\sin{r_{t0}}} \\
\sin{r_p} \cos{\alpha}
\end{split}
\end{equation}
\vspace{-5pt}
\begin{equation}
\Delta\theta = \arccos{\frac{\cos{r_p} - \cos{r_{t0}} \cos{r_{t1}}}{\sin{r_{t0}} \sin{r_{t1}}}}
\end{equation}
\vspace{-5pt}
\begin{equation}
\gamma'_{t1} = \arccos{\frac{\cos{r_{t0}} - \cos{r_p} \cos{r_{t1}}}{\sin{r_p} \sin{r_{t1}}}}
\end{equation}
\vspace{-5pt}

If $K < 0$, then:
\vspace{-5pt}
\begin{equation}
\begin{split}
r_{t1} = \arccos{\cosh{r_{t0}} \cosh{r_p}+\sinh{r_{t0}}}\\
\sinh{r_p} \cos{\alpha}
\end{split}
\end{equation}
\vspace{-5pt}
\begin{equation}
\Delta\theta = \arccos{\frac{\cosh{r_{t0}} \cosh{r_{t1}} - \cosh{r_p}}{\sinh{r_{t0}} \sinh{r_{t1}}}}
\end{equation}
\vspace{-5pt}
\begin{equation}
\gamma'_{t1} = \arccos{\frac{\cosh{r_p} \cosh{r_{t1}} - \cosh{r_{t0}}}{\sinh{r_p} \sinh{r_{t1}}}}
\end{equation}
\vspace{-5pt}

$\alpha = \beta_{t0} - \gamma_{t0}$. Angle $\alpha$ is the difference between object rotation and its geodesic of movement (C''$_0$C''$_1$), which has to stay constant if the object is not rotating over time. Hence:
\vspace{-5pt}
\begin{equation}
\beta_{t1} = \gamma_{t1} + \alpha
\end{equation}
\vspace{-5pt}

$\gamma'_{t1}$ and $\gamma_{t1}$ are supplementary angles, so:
\vspace{-5pt}
\begin{equation}
\gamma_{t1} = \Pi - \gamma'_{t1}
\end{equation}
\vspace{-5pt}

To find the $\theta$ coordinate, either subtract or add $\Delta\theta$ to the $\theta_c$ depending on whether the angle $\alpha$ or its explementary angle is used for the subsequent calculation.
\end{corollary}

\section{Results}
\subsection{Engine}

Using the method described in the previous section and OpenGL, we have created an engine that has a capability to calculate the objects and render the vector graphics in a non-Euclidean space with constant arbitrary curvature in the range of $-1\leq K \leq 1$.

The movement of an object can be shown dynamically by creating several time-lapse images collated from multiple screenshots of the game screen one over the other. These are shown in the figures 14 and 15 below. They show movement through different geodesics at $K=1$ and $K=-1$ on the left and right of each figure respectively.

\end{multicols}

\begin{figure}[H]
\vspace{-15pt}
\centering
\begin{subfigure}{.5\textwidth}
  \centering
  \includegraphics[width=.95\textwidth]{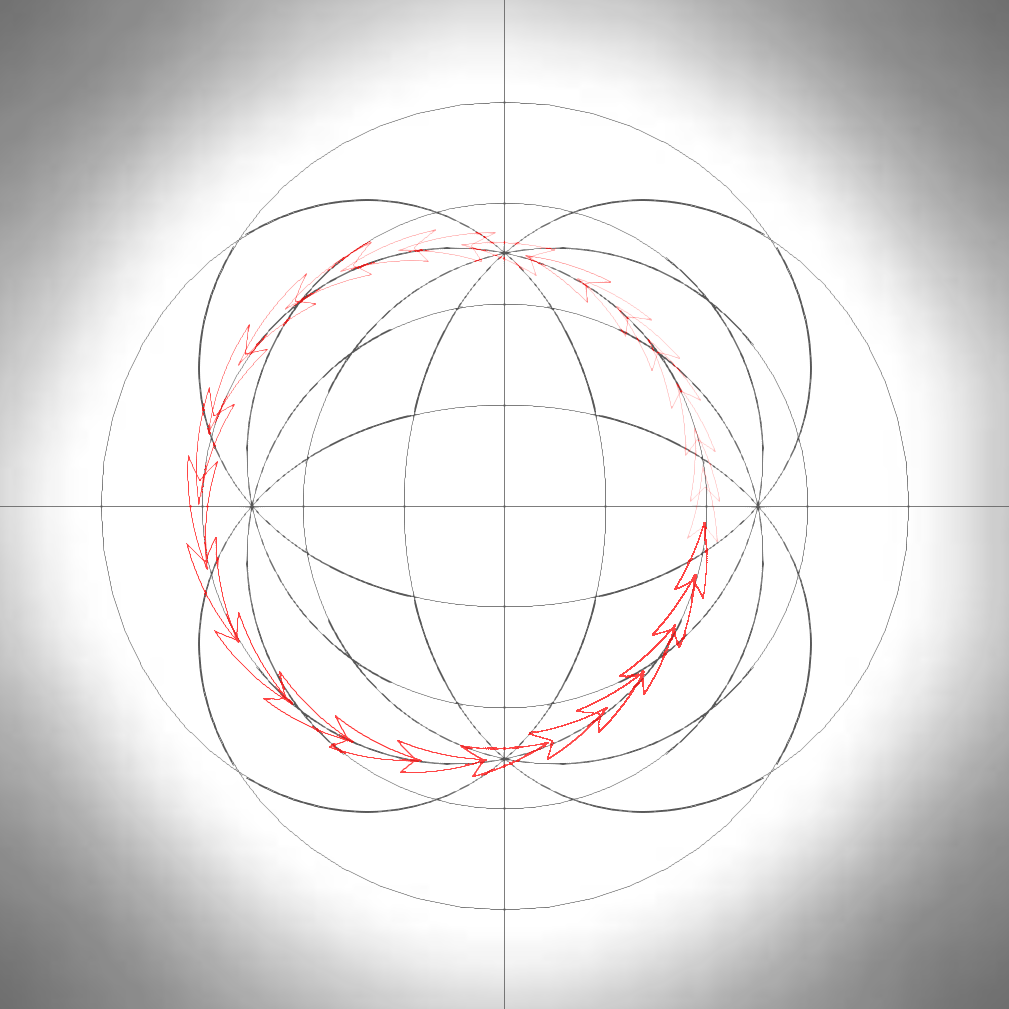}
  \caption{Spherical space ($K=1$)}
  \label{fig:SphericalMovement}
\end{subfigure}%
\begin{subfigure}{.5\textwidth}
  \centering
  \includegraphics[width=.95\textwidth]{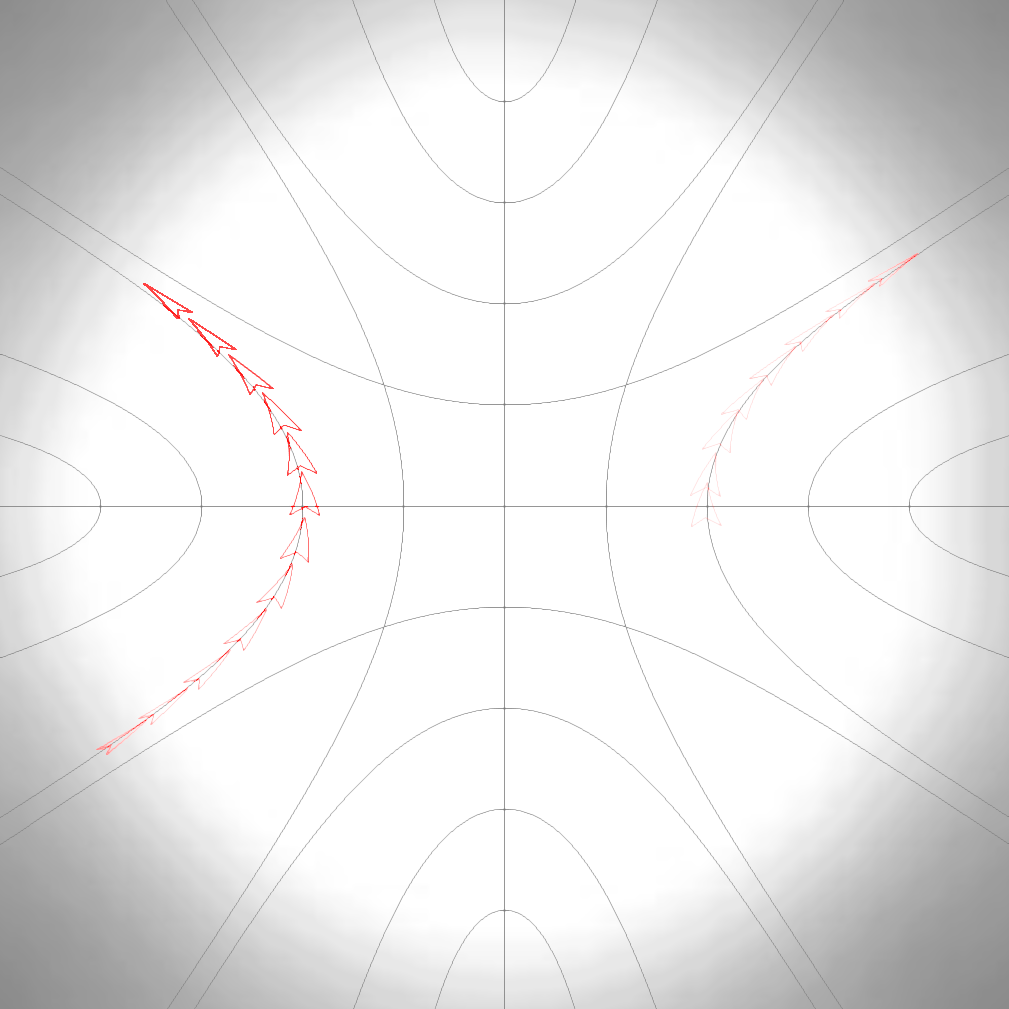}
  \caption{Hyperbolic space ($K=-1$)}
  \label{fig:HyperbolicMovement}
\end{subfigure}
\raggedright
\vspace{-5pt}
\caption{Movement along the geodesic orthogonal to the $0^o$ theta vector}
\vspace{-5pt}
\end{figure}

\begin{figure}[H]
\vspace{-15pt}
\centering
\begin{subfigure}{.5\textwidth}
  \centering
  \includegraphics[width=.95\textwidth]{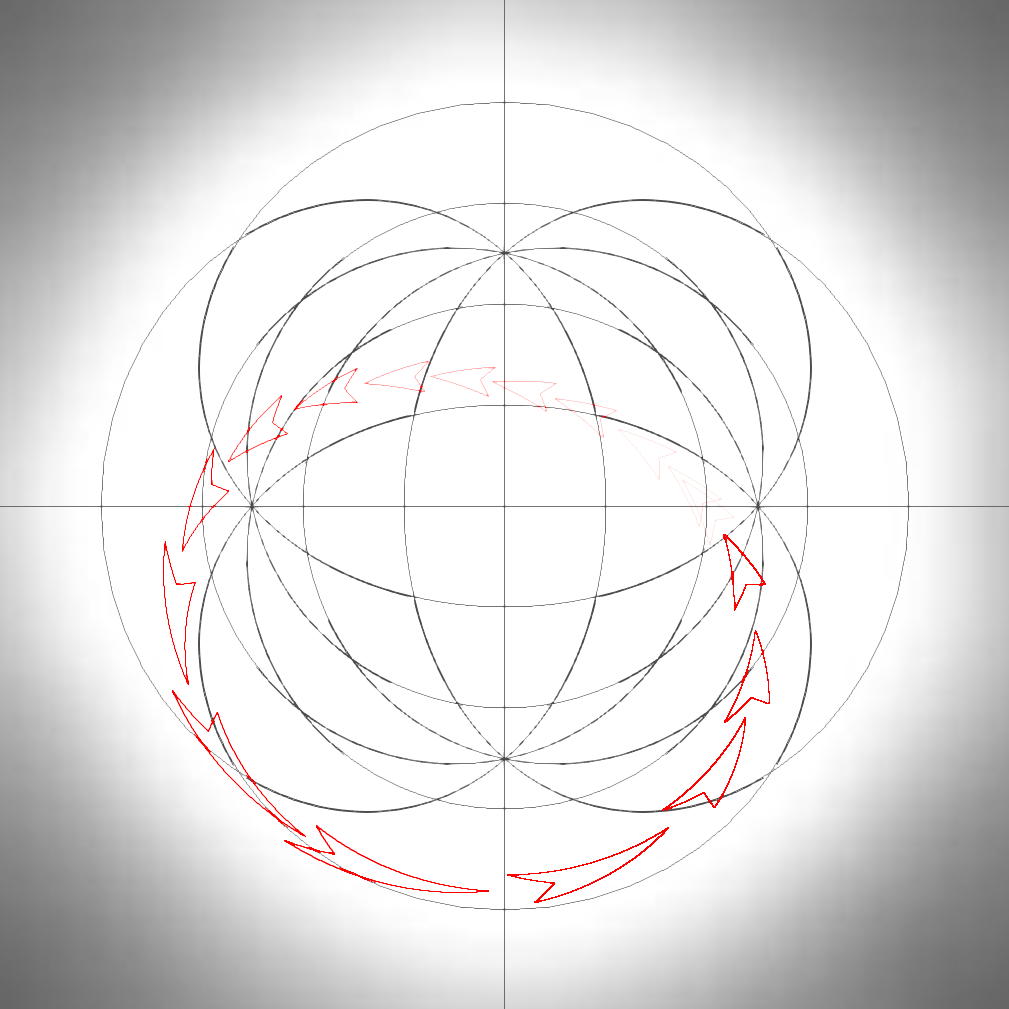}
  \caption{Spherical space ($K=1$)}
  \label{fig:SphericalMovement}
\end{subfigure}%
\begin{subfigure}{.5\textwidth}
  \centering
  \includegraphics[width=.95\textwidth]{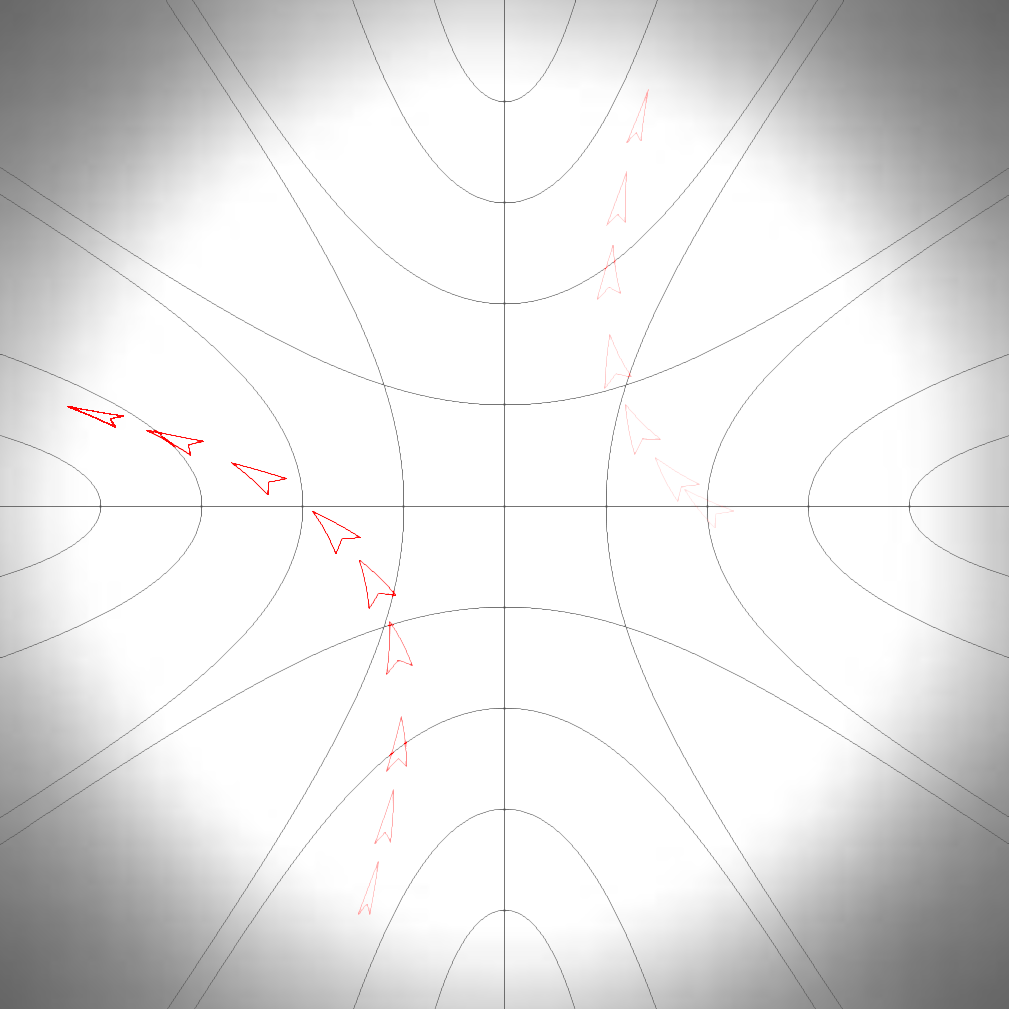}
  \caption{Hyperbolic space ($K=-1$)}
  \label{fig:HyperbolicMovement}
\end{subfigure}
\raggedright
\vspace{-5pt}
\caption{Movement along the geodesic at $45^o$ to the $0^o$ theta vector}
\vspace{-10pt}
\end{figure}

\begin{multicols}{2}

While these timeline images have the same starting position of the object, these have been used as an aid for visual comparison between images. The software can calculate the object flying in arbitrary direction with arbitrary speed as well as starting from arbitrary position in the space.

Another feature that was added to aid demonstration of the space is the cut-off of the worlds at a distance of $N$ pixels. This can be seen in the hyperbolic movement time-lapse image. While the hyperbolic space should be infinite, we chose to limit it in order to keep objects within the boundaries of the screen. This way they can be more easily observed. Because of the coordinate system used, it is easy to set or lift this limiting distance: the object's theta coordinate is increased by $\pi$ and then standardised. This makes it appear on the antipodal point of the limit circle with preserved velocity and orientation. This can be seen in Figure 14 (b) and Figure 15 (b). As the object crosses into the shaded area of the world, it is immediately reset on the antipodal point of the white limit circle.

\end{multicols}

\begin{figure}[H]
\vspace{-15pt}
\centering
\begin{subfigure}{.5\textwidth}
  \centering
  \includegraphics[width=.95\textwidth]{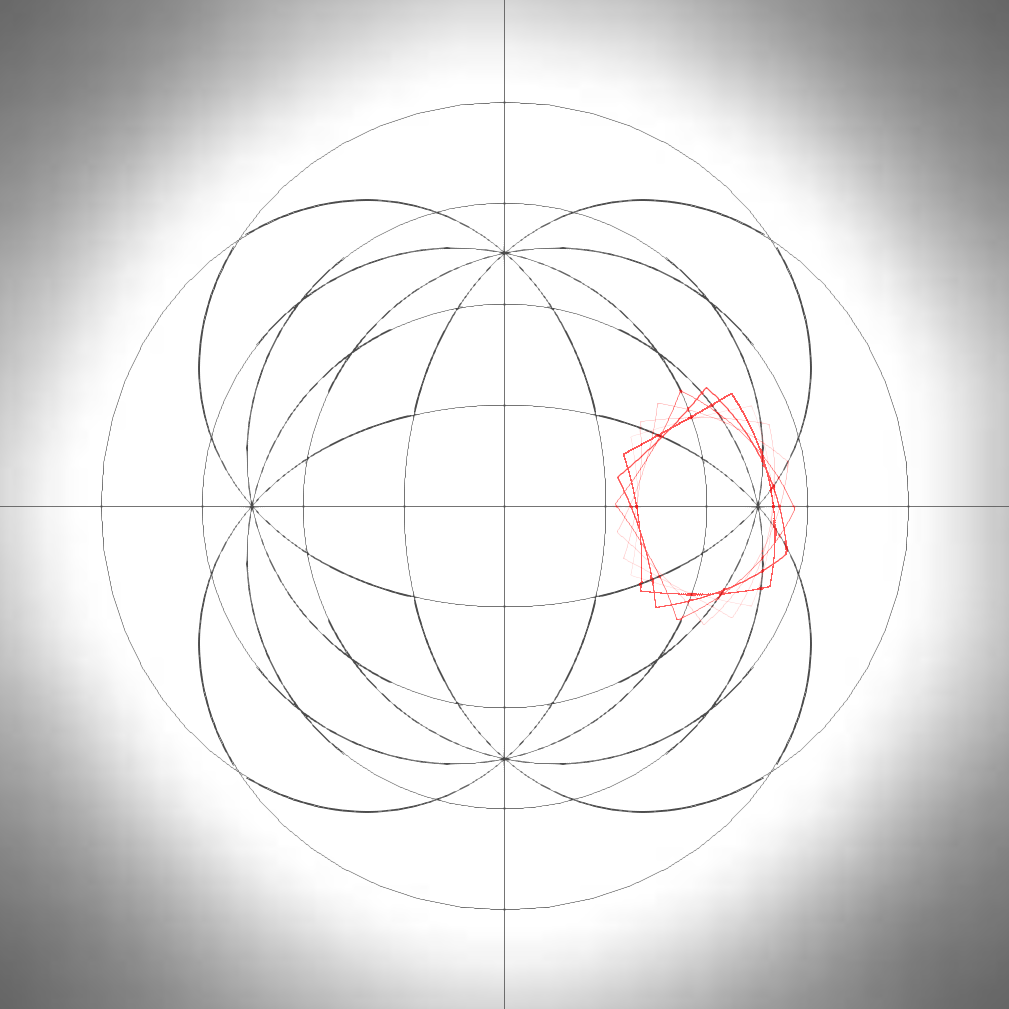}
  \caption{Spherical space ($K=1$)}
  \label{fig:SphericalMovement}
\end{subfigure}%
\begin{subfigure}{.5\textwidth}
  \centering
  \includegraphics[width=.95\textwidth]{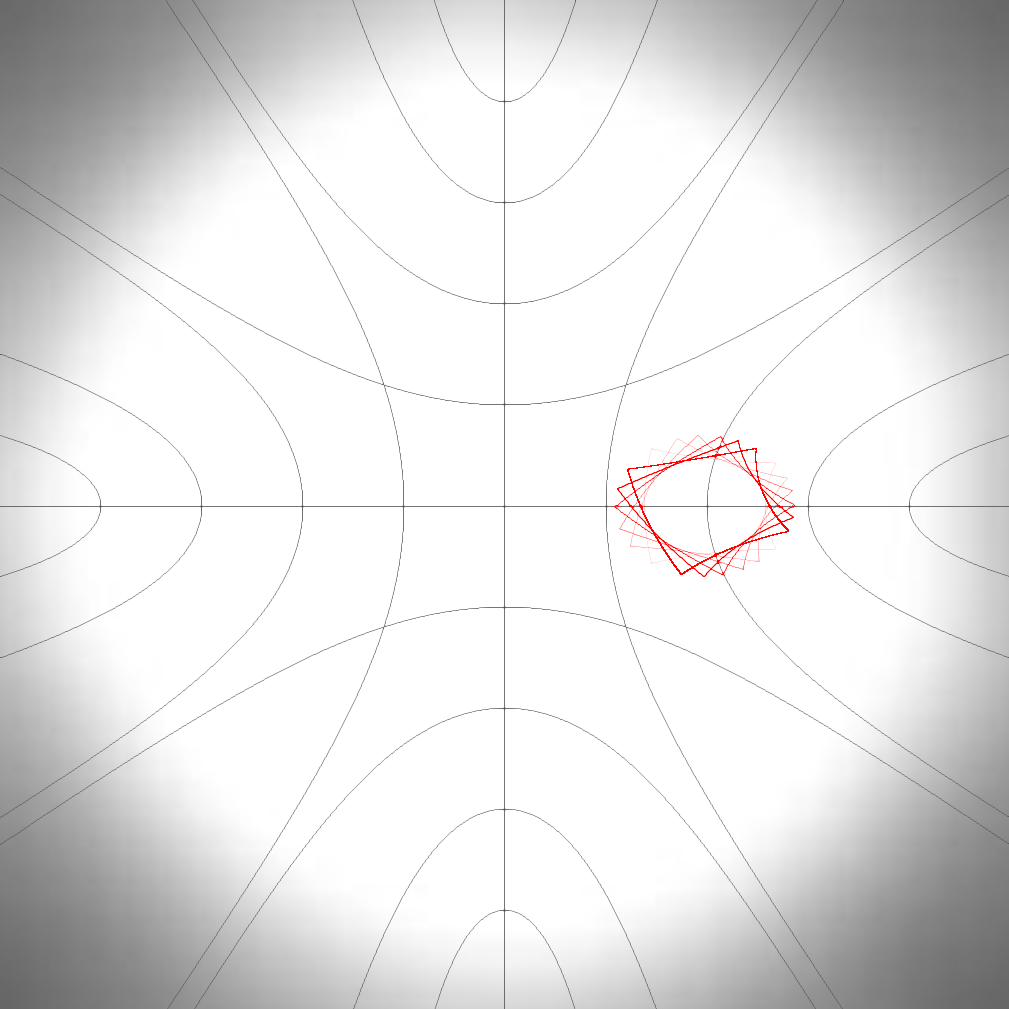}
  \caption{Hyperbolic space ($K=-1$)}
  \label{fig:HyperbolicMovement}
\end{subfigure}
\raggedright
\vspace{-5pt}
\caption{Rotation of the object around its centre point}
\vspace{-10pt}
\end{figure}

\begin{multicols}{2}
In addition to movement through curved space we decided to show a time-lapse image of a rotation of the `square' (in this case, the we describe a square by a quadrilateral that has 4 vertices equidistant from the centre of the object in local coordinates as well as being equally spaced out around the local reference point). Figure 16 shows a rotation of this object in spherical (a) and hyperbolic (b) space. Please note that the object is described with exactly the same parameters in hyperbolic and spherical spaces: $[(90,\frac{\pi}{8}), (90, \frac{3\pi}{8}), (90,\frac{5\pi}{8}), (90,\frac{7\pi}{8})]$

Note that the grid-lines in the time-lapse images have been created and rendered as separate game objects, hence there is no need to recalculate them manually when the curvature changes.

The curvature of the world can be changed in real time using keyboard inputs in a similar manner to controlling the object's acceleration and orientation. The curvature can be set to several pre-made values or changed continuously by adding or subtracting a small number from the current curvature variable every time the corresponding button is pressed.

\begin{figure}[H]
	\vspace{-5pt}
	\centering
	\includegraphics[width=.475\textwidth]{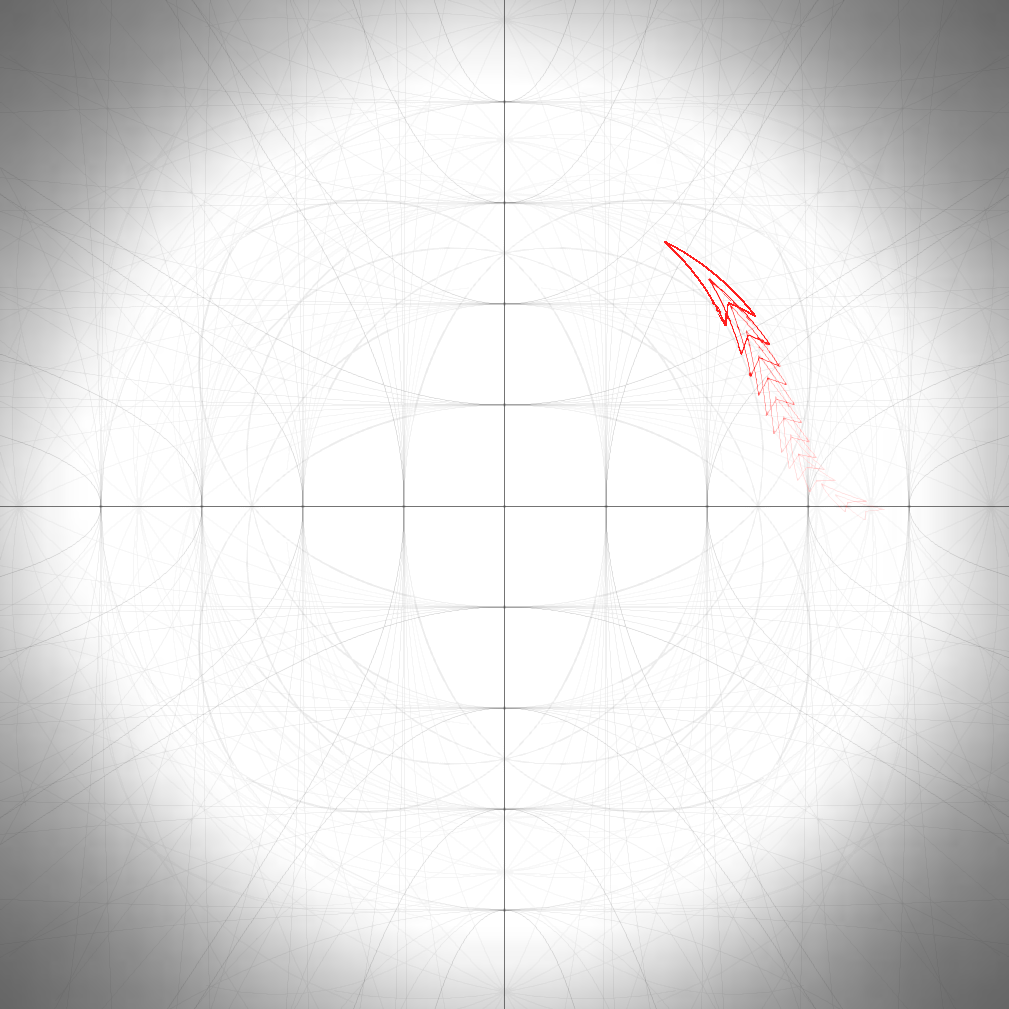}
	\caption{Movement of the object while changing the curvature from K=-1 to K=1}
	\label{fig:mixed}
	\vspace{-10pt}
\end{figure}

In the time-lapse image above (Figure 17) this feature is demonstrated by starting the object's movement in hyperbolic space and then gradually increasing the value of K in order to progress from a hyperbolic space through flat space and into a spherical space. This can be observed by not only the movement of the `spaceship' (i.e. red shape), but also the change in the way the grid lines are curved.

This feature allows for flexibility in the use of the engine. The gradual change in curvature can be used to explain the concept of curvature much more easily than the standard methods of imagining the spherical space or a flat projection of a sphere. 

We created a video \cite{youtube:2019} displaying the implementation.

\subsection{Complexity Analysis}

The complexity of the method can be calculated by examining its components individually and then combining to find the overall complexity.

Shape rendering is the more computationally and spatially expensive part of the method. In order to find all of the required points to display one shape correctly, positions of each vertex need to be calculated, which requires $O(v)$ time, where $v$ is the number of vertices. Subsequently, intermediate points on each edge have to also be computed, requiring $O(i)$ time to find all of the points on a single edge, where $i$ is the level of tessellation. So overall complexity to render the world with $s$ number of shapes would be $O(s*v*i)$. The best case would be equal to $O(n)$ complexity, if two of the terms are negligibly small (e.g. a world with small number of 2 vertex shapes). The worst case can be approximated to $O(n^3)$ if all terms were comparably large (e.g. rendering a world with a large number of shapes with many vertices, like circles).

Spatial complexity for shape rendering is only $O(v*i)$ as, while all of the vertices have to be stored in the memory between calculating and rendering a shape, these are then rewritten to store the next shape's data. So either $O(n)$ in the best case or $O(n^2)$ in the worst case.

Regarding object's movement calculation, only one calculation per object is required and the previous position record is overwritten, both spatial and time complexity is $O(n)$, where $n$ is the number of objects in the world that are being updated.

Additional cost comes from the use of trigonometric and hyperbolic functions in the calculations. These operations are slower to compute than simpler operations (the exact cost of these functions depends on the hardware being used). For example some systems use the AGM iteration \cite{brent:2010} method to compute elementary functions (including trigonometric functions), which is faster than the previously common Taylor series method.

\section{Discussion}

Overall the implementation of the method has good performance up to a certain number of objects or the amount of tessellation. For the first implementation we have been focussing on making sure the method was implemented correctly and was working continuously under any curvature in the range $-1 \leq K \leq 1$. This has been achieved and the next step is to build upon this implementation to achieve a more complete engine. There are several ways this method could be improved and expanded upon.

Our initial extension would be the implementation of parallelism. Calculating all the points in order to render the objects creates a bottleneck issue. This could be parallelised to achieve a time complexity of the order $O(n)$, where $n$ is the number of objects in the world. Performing some of the calculations directly on the GPU would be one of such solutions.

With regards to reducing the additional computational cost of trigonometric functions, several approaches can be taken here. One of the commonly used methods would be to use lookup tables: pre-calculating a table of values for all of the angles in order to just lookup the result instead of performing the calculation when called. This is a widely used approach, for example, Frank Rochet's implementation \cite{rochet:2004} completely replaces the standard C++ trigonometry functions with custom functions that use lookup tables. Another way to reduce the cost of the trigonometric functions would be to replace some of the calculations that include them with alternative approaches.

There are also ways to expand on this method in order for this engine to perform different operations. These could range from a collision detection system to Artificial intelligence algorithms, and from texturing of the objects (instead of simple vector graphics) to having a non-uniform curvature in the world. These are some of the improvements we are planning to implement to make a more complete graphics and physics engine.

There are multiple potential applications for this engine. The obvious one is using it for educational purposes, due to this representation of non-Euclidean space being more intuitive than standard projections used: Poincare disk and Upper Half-Plane models, as such  it could be a good introduction to the concepts on non-Euclidean geometry. 

A different use for the engine could be in cartography. \cite{gartner:2015} The engine could be modified to efficiently convert the cartographic data into different projections beyond just Azimuthal Equidistant. The dynamic modification of curvature could also be utilised as a zooming feature from a global to a more localised map. There is also research being done on the use of non-Euclidean geometry in ecology \cite{sutherland:2014} and climatology \cite{frei:2013}, where this engine could help with modelling dynamic systems in non-Euclidean environment.

Another use for the planned 3D engine could be in Astrophysics and Cosmology, specifically in modelling the different dynamic systems of cosmological objects. For this, the engine would have to be modified in order to have a capability of calculating non-uniform curvature of space-time. \cite{ross:2001} \cite{kragh:2012} But once completed, the ability to dynamically change the global curvature would provide an opportunity to test how the objects would be expected to behave in positively or negatively curved space-time.

\section{Conclusion}
We were aiming to create a physics and graphics engine capable of calculating the shapes, positions and movement of objects in a 2D space with constant positive or negative curvature; and then render these objects onto the screen using vector graphics. Another feature we were aiming to implement was the ability to change the curvature in real-time while the application was running. 

We have achieved both of these objectives by creating a method of calculating all of the points necessary to render the objects. This method is using spherical and hyperbolic trigonometry as well as a polar coordinate system to calculate the motion of the object along a geodesic through curved space and subsequently to find the positions of every vertex of the object given its position and shape in local coordinates. After that intermediate points are found in-between the vertices in order to render the curved lines onto the screen using. The additional objective of having a real-time modification of the curvature was achieved by having a dynamic recalculation of the objects at every step of the application execution. This method is currently capable of supporting multiple objects as well as a real-time control of one of the objects using the keyboard inputs, which can be seen in Figure 1. Currently all of the calculations are done sequentially, which creates a bottleneck that can slow down the frame rate of the application if there are a large number of objects that need to be rendered. This is not a critical problem, because there are several solutions that we are planning to research, such as parallelisation of the calculations.

The ability to change the curvature dynamically allows for a more intuitive representation of the non-Euclidean space, because the observer can see how the object changes in real-time as the curvature is increased or decreased. This also bridges the gap between the study of spherical and hyperbolic geometries, as both are rendered using the same method with the same projection. Hence the possibility of its application in multiple fields: cartography, education, physics, among others.
\end{multicols}
\newpage

\bibliographystyle{ieeetr}
\bibliography{mybib}

\end{document}